\documentstyle[aps,multicol]{revtex}
\input epsf.sty
\begin{document} 
\draft \title{Correlation of Tunneling Spectra in $\rm{\mathbf 
Bi_2Sr_2CaCu_2O_{8+\delta}}$ with the Resonance Spin Excitation} 
\author{J.\ F.\ Zasadzinski,$^{1,2}$ L.\ Ozyuzer,$^{1,2}$ N.\ Miyakawa,$^{3}$ 
K.\ E.\ Gray,$^{2}$ D.\ G.\ Hinks,$^{2}$ and C.\ Kendziora$^{4}$}
\address{$^{1}$Physics Division, Illinois Institute of Technology, Chicago, 
IL 60616, USA\\
$^{2}$Materials Science Division, Argonne National Laboratory, Argonne, 
IL 60439, USA\\
$^{3}$Department of Applied Physics, 
Science University of Tokyo, Tokyo, Japan\\
$^{4}$Naval Research 
Laboratory, Washington, D.C., 20375, USA.}
\date{\today}
\maketitle

\begin{abstract}
New break--junction tunneling data are reported in $\rm 
Bi_2Sr_2CaCu_2O_{8+\delta}$ over a wide range of hole concentration from 
underdoped ($T_c = 74$ K) to optimal doped ($T_c = 95$ K) to overdoped ($T_c = 
48$ K).  The conductances exhibit sharp dips at a voltage, $\Omega /e$, 
measured with respect to the superconducting gap.  Clear trends are 
found such that the dip strength is maximum at optimal doping and that 
$\Omega$ scales as $4.9 kT_c$ over the entire doping range.  These 
features link the dip to the resonance spin excitation and suggest 
quasiparticle interactions with this mode are important for 
superconductivity.
\end{abstract}
\begin{multicols}{2}

Acceptance of the phonon mechanism for electron pairing in 
conventional superconductors began with the isotope effect, showing $T_c$ 
scaling with a characteristic phonon energy, $\Omega_{\rm ph}$, and 
culminated with detailed, quantitative agreement between features in 
tunneling spectroscopy and the phonon spectrum measured by neutron 
scattering \cite{1}.  For high $T_c$ cuprates, however, tunneling and 
other spectroscopies have not led to a consensus on the pairing 
mechanism and in fact are currently being interpreted within radically 
different theoretical frameworks.  The doping dependence of spectral 
features can be a key in determining their origin.  Thus we present 
superconductor--insulator--superconductor (SIS) tunneling data on $\rm 
Bi_2Sr_2CaCu_2O_{8+\delta}$ over a very wide doping range from underdoped 
($T_c = 74$ K) to optimal doped ($T_c = 95$ K) to overdoped ($T_c = 
48$ K).  From zero bias up to the superconducting gap voltage, 
$2\Delta /e$, the measured conductances are close to that expected 
from the density of states found in mean--field models 
of a $d$--wave order parameter.  However, the conductances also reveal 
sharp dips at a voltage, $\Omega /e$, beyond the gap edge.  This dip 
feature is similar to structures ascribed to phonons in conventional 
superconductors and suggests that electrons are coupled to some type 
of collective excitation of energy $\sim\Omega$, measurable by the dip 
minimum.  Most importantly, the scaling of $\Omega$ as $4.9 kT_c$ over 
the entire doping range is close to that of the resonance spin 
excitation energy, $\Omega_{\rm res}$, found in neutron 
scattering \cite{2,3}. Thus the tunneling and neutron measurements, coupled 
together, present a strong case that spin excitations play an 
important role in the pairing mechanism of high $T_c$ superconductors.  

An unusual spectral dip feature in the tunneling data of 
$\rm Bi_2Sr_2CaCu_2O_{8+\delta}$ (Bi2212) was pointed out as early as 1989 in 
superconductor--insulator--normal metal (SIN) junctions \cite{4} where the 
dynamic conductance, $\sigma(V)$, is expected to be proportional to the 
electronic density of states, $N(E)$, with $E=eV$.  This feature was 
repeatedly observed in many subsequent tunneling studies, including 
scanning tunneling spectroscopy (STS) \cite{5,6}, break junctions 
\cite{5,7} and recently in a stack of intrinsic \textbf{c}--axis 
junctions \cite{8} within a Bi2212 crystal intercalated with HgBr$_2$.  
These consistent observations (e.g.  location and magnitude of the 
dip) in such a variety of junction types tends to rule out extrinsic surface 
related phenomena as being responsible \cite{1}. An earlier tunneling study showed that the dip 
represented a suppression of the density of states and that the
strength of the dip correlated with $T_c$ in overdoped Bi2212 \cite{5}. 
It was argued to be some form of strong coupling 
effect analogous to the dip features from the electron--phonon 
interaction in conventional superconductors \cite{5}. Phonon modes with 
energy $\Omega_{\rm ph}$ produce a dip feature with a minimum near 
$\Delta_s+\Omega_{\rm ph}$ in SIN junctions \cite{1} and near 
$2\Delta_s+\Omega_{\rm ph}$ in SIS junctions where $\Delta_s$ is the 
$s$--wave superconducting gap.  Although preliminary tunneling 
measurements on Bi2212 suggested that the dip voltage scaled with the 
maximum $d$--wave gap \cite{5,9,10}, $\Delta$, we will demonstrate 
here that the sharp dips observed in these new SIS junctions have a 
minimum at $2\Delta+\Omega$, where $\Omega$ is proportional to $T_c$.

The linking of the tunneling dip feature to the resonance spin 
excitation has been made possible due to the confluence of a number of 
experimental and theoretical developments.  Neutron scattering has 
shown that the resonance mode is generic to bi--layer high $T_c$ cuprates 
reaching a maximum energy of $41-43$ meV at optimal doping and tracking 
$T_c$ with underdoping \cite{3} or slight overdoping \cite{2}.  Other 
spectroscopies including angle resolved photoemission (ARPES) \cite{11,12} 
and optical conductivity \cite{13} have exhibited spectral features that 
can be explained by assuming the conduction electrons are coupled to 
the spin excitations.  Theoretical spin--fermion type models \cite{11,14} 
which have been invoked to explain the ARPES data have also been shown 
to produce dip features in the quasiparticle density of states and SIS 
tunneling spectra \cite{15,16} that resemble those reported
\begin{figure}
\centerline{
\begin{minipage}{1\linewidth}
\vskip-.78in
\centerline{\epsfxsize=1\linewidth \epsfbox{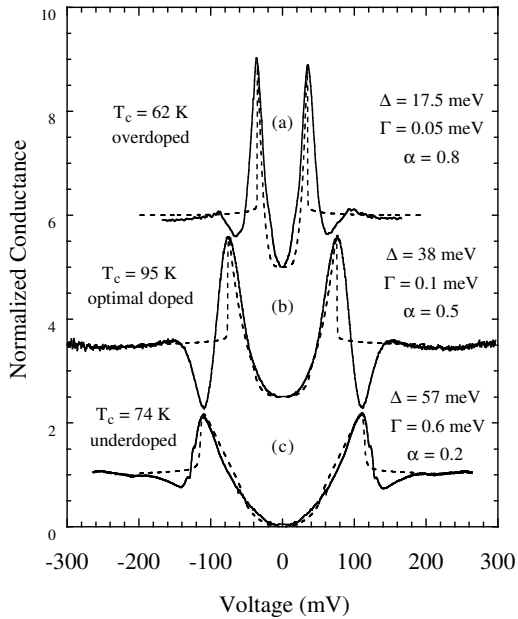}} \vskip-.28in
\caption{SIS tunneling conductances for a) overdoped, b) optimally 
doped and c) underdoped Bi2212.  Data have been normalized by a smooth 
background and shifted for clarity.  Dashed lines are BCS $d$--wave fits 
using $\Delta(\phi) = \Delta\cos 2\phi$, a scattering rate, $\Gamma$, and a 
weighting function, $f(\phi) = 1+{\alpha\cos}4\phi$ as described in 
ref.\ \protect\cite{14}.}
\label{fig1}
\end{minipage}}
\end{figure}
\noindent here. Unique to this tunneling work, however, is that we have measured 
the relations between $\Omega$, $T_c$ and $\Delta$ where the latter varies 
by a factor of six over this doping range.  Such a systematic study 
allows trends to be observed and thereby makes a more convincing case that the tunneling dip is due to the 
resonance mode and not some other excitations such as phonons. In 
addition, we demonstrate an interesting relation between the mode energy and the superconducting gap
that has provided some important insights into the nature of the resonance mode 
itself. The focus here is on SIS junctions because the dip is very pronounced in 
this geometry. The break junctions were obtained by a point contact technique 
(described elsewhere \cite{5,10}) on Bi2212 crystals oxygen doped over a 
wide range of hole concentration. We stress that these break 
junctions are formed under high vacuum, cryogenic conditions and they 
occur deep in the Bi2212 crystal \cite{17}, thereby minimizing surface 
contamination.  In Fig.\ \ref{fig1} the dynamic conductance spectra at 4.2 K 
are shown for three SIS break junctions on Bi2212: a) an overdoped, b) 
an optimally doped and c) an underdoped crystal.  These SIS spectra 
are a compendium of a much larger data set and they capture the 
principal features of interest.  The main conductance peaks reveal the 
energy gap at $|eV_p|=2\Delta$, which increases in the underdoped region even 
as $T_c$ decreases \cite{10}.  For $|eV|$ beyond $2\Delta$ there is a 
pronounced dip feature that is strongest at optimal doping.  The new 
optimal doped data of Fig.\ \ref{fig1} are quite similar to those of 
ref.\ \cite{10}, but exhibit a somewhat
\begin{figure}
\centerline{
\begin{minipage}{1\linewidth}
\vskip-.66in
\centerline{\epsfxsize=1\linewidth \epsfbox{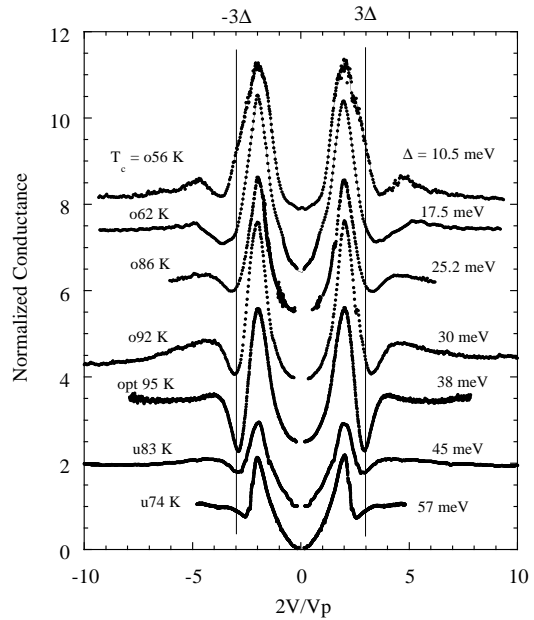}} \vskip-.4in
\caption{SIS 
tunneling conductances of Bi2212 for various hole dopings.  Notation 
is o$=$overdoped, opt$=$optimal doped and u$=$underdoped.  Voltage 
axis has been rescaled in units of $\Delta$. }
\label{fig2}
\end{minipage}}
\end{figure}
\noindent  stronger dip.  The spectra are compared with a  weak 
coupling, $d$--wave fit (see ref.\ \cite{17} for details) which includes a 
quasiparticle scattering rate, $\Gamma$.  Considering the simplicity of the 
$d$--wave model, the agreement in the sub--gap region (up to $eV=2\Delta$) is 
surprisingly good.  Even better agreement is found for SIN junctions where the 
$d$--wave 
fit captures the cusp feature found at zero bias. \cite{17}    

For $|eV|>2\Delta$ there is an immediate positive deviation from the fit, 
followed by the strong negative deviation 
(dip) and finally a recovery toward a hump feature.  There is 
no way to broaden the peaks in the $d$--wave model (e.g.\ by 
increasing the scattering rate, $\Gamma$) without severely reducing 
the peak heights and that would be clearly incompatible with the data.  
We emphasize that the identical behavior is found throughout the 
doping range and that these deviations from the $d$--wave fit resemble 
the strong coupling effects from the electron--phonon interaction in 
conventional superconductors \cite{1} suggesting that the dip features 
are due to some type of bosonic collective excitation (or a relatively 
narrow spectrum of excitations).

A larger set of spectra over the entire range of doping is plotted in 
Fig.\ \ref{fig2} on a voltage axis which is scaled by $\Delta=eV_p/2$.  In 
addition to the dip strength being maximum at the highest $T_c$, which 
suggests a strong coupling effect \cite{5}, another trend is evident.  On 
this plot it can be seen that while the dip minimum is very close to 
$3\Delta$ at optimal doping (as noted in previous work \cite{9,10}) it is 
significantly greater than (less than) $3\Delta$ with overdoping 
(underdoping).  This trend is easy to see here because the newer SIS 
spectra all have a peak height to background ratio $>2$ and exhibit 
sharp dip features.  Both of these trends clearly link the dip to the 
superconductivity, but rule out trivial connections such as proximity 
effects \cite{18}.  Also, the shift of $\Omega$ inferred from the dip 
minima in Fig.\ \ref{fig2} is consistent with theoretical predictions 
\cite{14} for the location of a resonance mode energy $\Omega_{\rm res}$ 
within the gap $2\Delta$ in the spin excitation spectrum.  Thus Figs.\ 
\ref{fig1} and \ref{fig2} make a strong case that: (1) the dip feature 
arises from quasiparticles coupled to some type of collective 
excitation of energy, $\Omega$, and (2) the doping dependence of 
$\Omega$ bears a qualitative resemblance to the resonance spin 
excitation in experiment \cite{2,3} and theory \cite{14}.

So far we have implied that the dip minimum might provide a quantitative 
measure of $\Omega$, but some justification for this is required, especially 
in the absence of a full microscopic theory.  Considering conventional 
phonon structures in $s$--wave superconductors \cite{1}, the dip 
minimum in SIS spectra would give a slight overestimate of the 
excitation energy.  However the cuprates are $d$--wave superconductors 
and the presence of gap nodes can affect the location of strong 
coupling features.  This result comes from the analysis of ARPES data 
in Bi2212 \cite{11,12} which show a similar dip/hump feature for 
electrons near the $(\pi,0)$ point i.e., maximum gap region.  The 
ARPES data can be analyzed within a model whereby the electrons near 
$(\pi,0)$ are interacting with a collective mode and the mode energy 
is given by the dip minimum \cite{11} which is at $\Delta+\Omega$.  
This model for Bi2212 has been extended \cite{15} to calculations of 
the SIS spectra by considering a $d$--wave gap, a realistic Fermi 
surface and the coupling of electrons of all momenta to the resonance 
spin excitation at the $(\pi,\pi)$ point.  The result is that the dip 
minimum in SIS junctions is very close to $2\Delta+\Omega$.  It should 
be noted that a dip feature has been observed in the angle--integrated 
photoemission spectra of Pb and Nb and linked to peaks in the phonon 
density of states \cite{19} clearly demonstrating that this type of 
dip in photoemission can be a characteristic of strong coupling 
superconductivity.

\begin{figure}
\centerline{
\begin{minipage}{1\linewidth}
\vskip-.9in
\centerline{\hskip-.4in\epsfxsize=1\linewidth \epsfbox{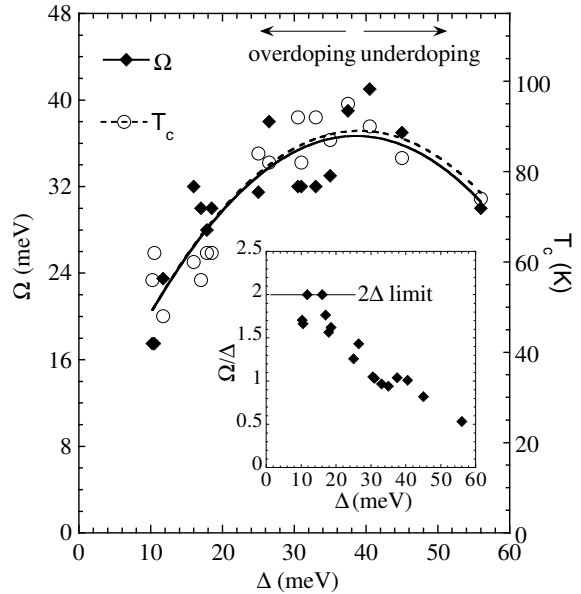}}\vskip-.4in
\caption{Measured mode energy $\Omega$ and bulk $T_c$ value 
vs.\ measured gap value $\Delta$ for 17 junctions over a wide doping 
range from u74K to o48K.  Solid and dashed lines are quadratic fits of 
$\Omega$ and $T_c$ vs.\ $\Delta$.  Inset shows $\Omega /\Delta$ vs.\ 
$\Delta$.}
\label{fig3}
\end{minipage}}
\end{figure}

Considering the above discussion, it seems reasonable to extract $\Omega$ 
by assuming the dip minimum is at $2\Delta+\Omega$.  In Fig.\ \ref{fig3} are 
plotted the measured values of $\Omega$ and $T_c$ vs.\ the measured 
$\Delta$ obtained on 17 SIS junctions from crystals with bulk $T_c$ ranging 
from 74 K (underdoped) to 95 K (optimal doped) to 48 K (overdoped).
It is found that the maximum measured value of $\Omega\sim 42$ meV occurs near 
optimal doping and agrees with the maximum 
resonance mode energy obtained in neutron scattering \cite{2,3}. The scatter 
in the data of Fig.\ \ref{fig3} arises from an experimental uncertainty in 
determining the mode energy, $\Omega$, from the dip minimum and also 
from the narrow spread of gap values from junctions made on different 
regions of a crystal with a common bulk $T_c$.  Despite the scatter in 
the data there is an unmistakeable correlation between $\Omega$ and 
$T_c$.  This is seen with simple quadratic fits to $\Omega$ and $T_c$ 
vs.\ $\Delta$ (solid and dashed line respectively in Fig.\ \ref{fig3}) 
which are nearly congruent and lead to the scaling $\Omega/kT_c\sim 
4.9$ which we note is in good quantitative agreement with
\noindent  neutron 
results \cite{2,3} for $\Omega_{\rm res}/kT_c\sim 5.1-5.5$. This scaling is 
reminiscent of the conventional isotope effect in superconductors 
where $T_c$ scales with $\Omega_{\rm ph}$.  We note that the tunneling $\Omega$ 
scales with $T_c$ far into the overdoped region, well beyond what has 
been measured so far for the neutron resonance.

In the plot of $\Omega/\Delta$ vs.\ $\Delta$ (inset of Fig.\ \ref{fig3}) the 
relation of $\Omega$ to $\Delta$ is more clearly seen, providing important 
information on the nature of the excitation probed by tunneling.  For the 
most overdoped region, the mode energy approaches but never exceeds 
$2\Delta$, and $\Omega /\Delta$ monotonically decreases as doping 
decreases and the superconducting gap grows.  Thus the excitation itself is 
linked to the superconducting gap which varies by a factor of six in this 
study.  This behavior would seem to rule out phonons completely and instead
suggests an electronic excitation. It is consistent with general ideas about collective modes of the conduction 
electrons in superconductors \cite{20} (i.e., that mode energies 
$>2\Delta$ are heavily damped), as well as with specific models of the 
resonance spin excitation \cite{14}. In the latter, the overdoped region is 
considered to have weaker coupling of electrons to spin excitations 
and therefore the resonance mode is close to the gap edge, $2\Delta$, in the 
spin excitation spectrum.  As doping decreases and the coupling gets 
stronger, the superconducting gap gets larger but the mode energy 
moves deeper into the gap and thus $\Omega /\Delta$ decreases.  In this 
picture the neutron resonance is a spin exciton \cite{14} and we believe 
the tunneling data provide the first evidence for such a viewpoint.      
 
To summarize, we have used the doping dependence of the tunneling dip  
feature, as shown in Fig.\ \ref{fig3}, to link it to the resonance spin excitation. 
Since the dip resembles a strong coupling effect, the tunneling data are 
also providing evidence that 
spin excitations are playing a crucial role in superconductivity.    It 
should be noted that a similar dip feature has been observed in the 
superconducting tunneling spectra of a heavy fermion superconductor 
\cite{21} which has also been linked to a peak that develops in the spin 
excitation spectrum.  Thus a spin fluctuation mechanism may have 
relevance to superconductors beyond the high $T_c$ cuprates.

It is important to contrast this interaction with phonon mediated pairing.  Phonons are collective excitations of the lattice 
and therefore exist in the normal state, whereas this collective spin 
excitation mode seems to develop only below $T_c$ (except for heavily underdoped 
materials).  Thus this scenario displays a remarkable feedback effect: 
superconductivity hollows out a gap in the spectrum of damped spin 
excitations, allowing a propagating collective mode to exist, but then 
that mode is at least partly responsible for the pairing mechanism.  
If, in addition, there are phonon contributions to the pairing, the phonon 
structures are not showing up in a clear and reproducible manner.
The selective coupling of electrons near $(\pi,0)$ to spin 
excitations near $(\pi,\pi)$ as suggested in ARPES \cite{11,12,14,15} leads 
naturally to $d$--wave superconductivity \cite{22} and explains the 
mysterious asymmetry of the dip strength with voltage polarity as 
observed in SIN tunneling \cite{4,5,6}. In this picture, the quasiparticles at 
$(\pi,0)$ are the most strongly renormalized and since this is an occupied 
state in Bi2212 \cite{11,12} the strongest dip features should be found for 
bias voltages which remove electrons, as is observed in SIN 
experiments and model calculations \cite{15}. There are still a number of 
unresolved issues.  As with phonon structures in conventional 
superconductors \cite{1}, one hopes to show that the tunneling spectra 
quantitatively reproduce the measured $\Delta$ and $T_c$.  This will require a 
microscopic model, and the present results suggest that a 
Fermi--liquid, Eliashberg type of approach will work.  While it is true 
that the pseudogap leads to a distinctly non--Fermi--liquid normal 
state, superconductivity seems to restore the quasiparticle concept.  
Finally, we should note that similar dip features have been observed 
in the tunneling spectra of $\rm Tl_2Ba_2CuO_6$ indicating that the neutron 
resonance ought to be observed in a cuprate with a single Cu--O layer 
per unit cell \cite{23}.

\acknowledgements{The authors benefitted considerably from discussions 
with A.\ Chubukov, B.\ Janko and M.\ Norman.  This work was partially 
supported by U.S.\  DOE--BES Contract No.  W--31--109--ENG--38, NSF--STCS 
Contract No.\  DMR 91--20000, and a Grant--in--Aid for Encouragement of 
Young Scientists from the Ministry of Education, Science and Culture, 
Japan.}

\end{multicols}

\begin{references}
\bibitem{18} Proximity effect models (as 
described in ref.  \cite{1}) cannot account for the trends observed.  
Furthermore they ascribe the dip/hump feature to the bulk 
superconducting gap which overestimates the gap by a factor of two or 
three.
\end{references}
\end{document}